\documentclass[useAMS,usenatbib,usegraphicx]{mn2e}

\def\msun{M$_\odot$}
\def\zsun{Z$_\odot$}
\def\al{$\alpha\lambda$}
\def\tn{$\tau_{nx}$}

\title[Two populations of progenitors for type Ia SNe]
{Two populations of progenitors for type Ia supernovae?}

\author[F. Mannucci et al.]{
F. Mannucci$^1$\thanks{E-mail:filippo@arcetri.astro.it},
M. Della Valle$^{2,3}$, and
N. Panagia$^{4,5}$\\
$^1$INAF, Istituto di Radioastronomia, Largo E. Fermi 5, 50125 Firenze, Italy\\
$^2$INAF, Osservatorio Astrofisico di Arcetri, Largo E. Fermi 5, 50125
Firenze, Italy\\
$^3$Kavli Institute for Theoretical Physics, UC Santa Barbara, CA 93106, USA
$^4$STScI, 3700 San Martin Drive, 
   Baltimore, MD 21218, USA\\
$^5$Also, INAF, Rome, Italy; and Supernova Ltd., Virgin Gorda, BVI
}

\begin{document}

\date{Submitted 2005 August; accepted 2006 April}

\pagerange{\pageref{firstpage}--\pageref{lastpage}} \pubyear{2006}

\maketitle

\begin{abstract}
We use recent observations of the evolution of the type Ia Supernova (SN
Ia) rate with redshift \citep{dahlen04}, the dependence of the SN Ia
rate on the colours of the parent galaxies \citep{mannucci05}, and the
enhancement of the SN Ia rate in radio-loud early-type galaxies
\citep{dellavalle05} to derive on robust empirical grounds, the
distribution of the delay time (DTD) between the formation of the
progenitor star and its explosion as a SN.  Our analysis finds:
i) delay times as long as 3--4 Gyr, derived from observations of SNe Ia at
high redshift, cannot reproduce the dependence of the SN Ia
rate on the colors and on the radio-luminosity of the parent galaxies,
as observed in the local Universe;  
ii) the comparison between observed SN
rates and a grid of theoretical ``single-population'' DTDs shows that
only a few of them are possibly consistent with observations. 
The most successful models are all predicting a   
peak of SN explosions soon after star formation and an
extended tail in the DTD, and can reproduce the data but only
at a modest statistical confidence level;    
iii) present data are
best matched by a bimodal DTD, in which about
50\% of type Ia SNe (dubbed {\it ``prompt''} SN Ia) explode soon after
their stellar birth, in a time of the order of $10^8$ years, while the
remaining 50\% ({\it ``tardy"} SN Ia) have a much wider distribution,
well described by an exponential function with a decay time of about 3
Gyr. 

The presence in the DTD of both a strong peak at early times 
and a prolonged exponential tail, coupled with the well established
bimodal distribution of the decay rate ($\Delta m_{15}$) and the
systematic difference observed in the expansion velocities of the ejecta
of SNe Ia in Ellipticals and Spirals, suggests the existence of two
classes of progenitors. We discuss the cosmological implications of this
result and make simple predictions, which are testable with future
instrumentation.
\end{abstract}
%
\section{Introduction}
Type Ia supernovae (SNe) SNe are very important objects in modern
cosmology because they are bright sources that can be detected up to
large distances and it appears that their intrinsic luminosities
can be inferred directly from  their light curves. Exploiting these
properties, the study of SN Ia at high redshifts has allowed the
discovery of the cosmic acceleration
\citep{perlmutter98,riess98,perlmutter99}.  Even though these objects
are commonly believed to be associated with the explosion of a
degenerate star as a white dwarf (e.g. Hillebrandt \& Niemeyer, 2000)
the nature of the progenitors of this class of SNe is not firmly
established, and several explosion patterns are possible 
(see, e.g., Branch et al., 1995; Yungelson, 2004)
and each of these may dominate at different redshifts. 
As a consequence, the existence of systematics affecting SNe Ia at
different redshifts cannot be ruled out 
(e.g. Kobayashi et al., 1998; Nomoto et al., 2003)
and it is worth being further investigated for possible
cosmological implications.

Different explosion models 
\citep{greggio83,yungelson00,matteucci01,belczynski05,greggio05} predict
different delay times between the formation of the progenitor
system and the SN explosion.  Differences in the expected delay times
are  testable by the observations (Madau, Della Valle \& Panagia 1998,
Sadat et al. 1998, Dahlen \& Fransson 1999) so that constraining the
Delay Time Distribution (DTD) will permit one to ascertain the nature of
SN Ia progenitors by confirming or excluding some of these models.
%
\section{New observational evidence}
 
Recently, three important observational results were established:
\begin{enumerate}
 
\item The evolution of the SN Ia rate with redshift is now measured up
to z$\sim$1.6 \citep{hardin00,pain02,strolger03,madgwick03,cappellaro04,
galyam04,mannucci05,barris06}.  The results at the highest redshifts,
derived by the GOODS collaboration
\citep{dahlen04,strolger04,strolger05}  show that the SN rate rises
up to z$\sim$0.8, when the Universe was 6.5 Gyr old (see panel b of
Fig.~1), and decreases afterward. However,  the detection of SNe at
$z\sim1.5$ is very challenging for the current instrumentation,
resulting in a small  SN sample (2 SNe in the Dahlen et al. highest
redshift bin),  and in a very uncertain rate. 

This behavior can be compared with the cosmic Star Formation History
(SFH) which continues to rise up to z$\sim2.5$, i.e., at a time about 4
Gyr earlier  \citep{madau98b,giavalisco04}. These results have been
interpreted  by \citet{dahlen04} and \citet{strolger05} as evidence of a
very long delay time ($\sim3-4$ Gyr)  between the formation of the stars
in the binary system and  the explosion of a SN Ia.

\item Recently we have measured the SN rate per unit mass in the
local Universe \citep{mannucci05}, and we have found a very strong
dependence of the rate on the (B$-$K) colour of the parent galaxies: 
blue galaxies (latest Hubble types) exhibit  a SN Ia rate 
a factor of $\sim 30$ higher than that of red galaxies (early types).
This result indicates that the delay time must have a wide 
distribution.   In star forming galaxies, the delay time must be as
short as the timescale of colour evolution ($\sim 0.5$ Gyr),  while the
existence of supernovae in  galaxies without any recent star formation
argues that at least some SN Ia have long delay times. For this reason
\citet{mannucci05},  following earlier suggestions
\citep{dallaporta73,dellavalle94,panagia00}, have proposed, on robust
empirical grounds, the existence of two populations of progenitors,
one related  to the young stellar  population, with rates proportional
to the recent SFR, the other related to the old stars, with rates
proportional to the total stellar  mass.

\item Della Valle et al. (2005) demonstrated that early-type  radio-loud
galaxies show a strong enhancement, by a factor of about 4, of the SN Ia
rate with respect to the radio-quiet sample, with a constant rate
between the samples being rejected at a 99.96\% confidence level (see also
\citet{dellapan03}).  Both the radio activity and the SN rate
enhancement are interpreted in terms of episodes of star formation due
to merging with small galaxies.  Since each episode of radio activity is
estimated to lasts about $10^8$ years \citep{srinand98,wan00},
under this hypothesis the evolutionary time for most SN Ia in
radio-loud galaxies must also be around 100 million years.

\end{enumerate}

All the above issues constitute observational links between the epochs
of star formation and SN explosion, and, therefore, can be used to 
constrain the DTD over different timescales.  The evolution of the SN Ia
rate with cosmic time is sensitive to long timescales (a few Gyr). The 
dependence of the local rate with the parent galaxy colour samples
timescales of the order of the colour evolution of the galaxies, i.e.,
0.5-1 Gyr. The relation between SN rate and radio power gives
information on the timescales of the order of $10^8$ years,
corresponding to the radio activity lifetime.

\section{Predicting the SN rates}

The aim of this work is to identify, among the possible DTDs,
which ones can simultaneously match all these three results.   We will
test a set of possible DTDs spanning a broad range of possibilities,
using both simple analytical formulations and the results from
theoretical works. For each DTD we compute the corresponding expected
rates, and compare the results with the above mentioned observational
results.

We have assumed that all stars in the 3--8 \msun\ mass range are
possible progenitors of SNe Ia (e.g., Nomoto 1994), and have a
probability $\eta$ \citep{madau98} of actually exploding as such within
an Hubble time from formation.  \citet{madau98} have estimated this
``explosion efficiency'' $\eta$ to be of the order of 5-10\%. Similar
results have been obtained by \citet{dahlen04} and \citet{greggio05}. 
We would like to note here that assuming a different mass range 
for the mass of the primary would only change the inferred overall
explosion efficiency by a factor that depends on the adopted mass limits
and the initial mass function (IMF) (see section~\ref{sec:mass}) without
affecting the conclusions about the nature of the progenitors
themselves.

\subsection{The evolution with redshift}

The evolution of the cosmic SN rate with redshift was modeled by a
simple convolution of the assumed DTD with the observed cosmic SFH
corrected for dust extinction, as derived by  \citet{giavalisco04}. This
is based on a decade of observations at different wavelengths and with
various techniques, starting from the pioneering work of Steidel and
coworkers \citep{steidel96}.

The need for a large correction for dust extinction introduce an
important uncertainty. The SFH derived directly from UV observations  is
similar in shape to the corrected one, but is about a factor of 3 lower.
As a result, the use of  this SFH would produce  very similar results
but would imply a number of SNe per unit mass in stars three times
larger.

We adopted the functional form of the SFH as a function of the cosmic
time $t$ (expressed in Gyr) as given by \citet{strolger04}, i.e. :\\

$SFH(t) = a\left(t^be^{-t/c}+d~e^{d(t-t_o)/c}\right)$ ~~~\msun/yr/Mpc$^3$ \\

\noindent
where a=0.182, b=1.26, c=1.865, d=0.071 and $t_0$ is the current age of
the universe, 13.47 Gyr for the assumed cosmology
$(h,\Omega_M,\Omega_\Lambda)=(0.7,0.3,0.7)$.

To properly take into account the time elapsed between  star formation
and the creation of white dwarfs that eventually will explode, the DTDs
have been  convolved with a function of time that, for a given stellar
generation, gives the number of white dwarfs accumulating as a result of
stellar evolution.  Adopting a Salpeter Initial Mass Function (IMF),
half of the white dwarfs from stars in the 3--8 \msun\ mass range form
within the first 130 Myr.

For the comparison, we have used the observed rates in
\citet{dahlen04}, \citet{strolger03}  and \citet{mannucci05}.  The SN
rate per unit mass of this latter work was transformed to a SN Ia rate
per unit volume by using the local 2dF luminosity function
\citep{cole01}, obtaining $3.5\cdot10^{-5}$ SN Ia/yr/Mpc$^3$.

\subsection{The dependence on the colour of the parent galaxy}
\label{sec: color}

In \citet{mannucci05} we have already demonstrated that the dependence
of the rate on the colour of the parent galaxy can be reproduced by a
simplified model, which comprises two components, one proportional to
the SFR (about 40\% of the core-collapse SN rate)  and one
proportional to the total stellar mass.   (about 4.4$\cdot10^{-4}$
SNe per year per $10^{10}$\msun\ of stars). This is consistent with a
DTD having two components: a prompt one,  exploding with a time scale
comparable to that of the core-collapse (CC) SNe; and a second component
characterized by a long delay, so that the corresponding rate  depends
only on the integrated history of the parent galaxy. This result was
obtained by using solely observed properties of galaxies and SNe,  and,
therefore, does not depend on any model of galaxy formation.

Here, we adopt a more refined and accurate approach, and use galaxy
formation models to link their observed properties with their past
history.  We reproduce the properties of the galaxies by  calculating a
large number of model galaxies (see Table~\ref{tab:models}) with widely
different properties. We used  the \citet{bruzual03} models with
different SF histories (from single burst to a rate extended over one
full Hubble time),  metallicities from 2\% to 250\% solar,  multiple
bursts of star formation,  and ages between 30 Myr and 13 Gyr.  For each
model galaxy we compute the present-day (at z=0) (B$-$K) colour and the
present-day SN Ia rate, obtained by convolving the SFH of that galaxy
with the assumed DTD.

The dependence of the rate on the colours is obtained by subdividing the
galaxies into the same bins of colours, as done for the galaxy sample
monitored for SN discoveries as in Mannucci et al. (2005),  and
averaging the SN rate in each bin.

To compute a meaningful cosmic average to be compared with the
observations, we impose that the collection of galaxy models used
reproduces the main integrated properties of the observed  galaxy
population. This was done by assigning appropriate weights to each
galaxy model to best match both the observed distribution of (B$-$K)
colour of the galaxies in the local universe and the evolution of the
cosmic star formation rate. Many solutions are possible, corresponding
to different mixes of galaxy models.  The variance in these predictions
was therefore estimated by creating many different sets of galaxy
samples that, with different mixes of galaxy models,  can still 
satisfy the requirements on the (B$-$K) colours and the cosmic SFH
with different mixes of models and different weights.  In all cases the
peak-to-peak  uncertainty range, represented as a shaded area in 
the panels (c) of Figures~1--5, is small enough to conclude that these
predictions are robust with respect to details of the galaxy models.

For each model, we have computed the value of reduced
$\chi^2$, i.e. $\chi^2_r=\chi^2$/D.O.F. by fully considering the
Poisson nature of the errors, and comparing the observed and expected
numbers of the events. In the figures we report the associated 
probability P that the discrepancies are merely due to statistical
fluctuations, i.e., the probability that the expected rates are
consistent with the  observed values. The reported values of $\chi^2_r$
and P do not take into account the ``model'' uncertainties, shown as
shaded areas in panels (c) and (d), as they are of  systematic rather
then statistical nature.  
Nevertheless, except for the case discussed in section~\ref{sec:single}, 
all the models contributing to the shaded areas give agreements 
comparable to the reported central value.

\begin{table}
\caption{List of the main properties of the used galaxy models}
\label{tab:models}
\begin{tabular}{cll}
\hline
\hline
Model & SF history   & Metallicity \\
\hline
 mod1 & 0.1 Gyr long Burst                  & \zsun \\		
 mod2 & exp. decay with $\tau$=0.2 Gyr      & 40\% \zsun \\	
 mod3 & exp. decay with $\tau$=0.2 Gyr      & \zsun \\		
 mod4 & exp. decay with $\tau$=0.2 Gyr      & 250\% \zsun \\	
 mod5 & exp. decay with $\tau$=5.0 Gyr      & 2\% \zsun \\	
 mod6 & exp. decay with $\tau$=5.0 Gyr      & 20\% \zsun \\	
 mod7 & exp. decay with $\tau$=5.0 Gyr      & 40\% \zsun \\	
 mod8 & exp. decay with $\tau$=5.0 Gyr      & \zsun \\		
 mod9 & exp. decay with $\tau$=8.0 Gyr      & 20\% \zsun \\	
mod10 & exp. decay with $\tau$=8.0 Gyr      & 40\% \zsun \\	
mod11 & exp. decay with $\tau$=8.0 Gyr      & \zsun \\		
mod12 & exp. decay with $\tau$=8.0 Gyr      & 250\% \zsun \\	
mod13 & 50\% mod3 + 50\%  mod1              & \zsun \\		
mod14 & 90\% mod3 + 10\%  mod1              & \zsun \\		
mod15 & 97\% mod3 + 3\%   mod1              & \zsun \\		
mod16 & 99\% mod3 + 1\%   mod1              & \zsun \\		
mod17 & 50\% mod3 + 50\%  mod1 after 3 Gyr  & \zsun \\		
mod18 & 90\% mod3 + 10\%  mod1 after 3 Gyr  & \zsun \\		
mod19 & 97\% mod3 +  3\%  mod1 after 3 Gyr  & \zsun \\		
mod20 & 99\% mod3 +  1\%  mod1 after 3 Gyr  & \zsun \\		
mod21 & 50\% mod3 + 50\%  mod1 after 10 Gyr & \zsun \\		
mod22 & 90\% mod3 + 10\%  mod1 after 10 Gyr & \zsun \\		
mod23 & 97\% mod3 +  3\%  mod1 after 10 Gyr & \zsun \\		
mod24 & 99\% mod3 +  1\%  mod1 after 10 Gyr & \zsun \\		
\hline
\end{tabular}
\end{table}

\subsection{The dependence on the radio power of the parent galaxy}

\citet{dellapan03} and \citet{dellavalle05} have demonstrated that the
SN Ia rate is higher in radio-loud Ellipticals than it is in radio-quiet
ones (about a factor of 4). These authors proposed that both the radio
activity and the excess in SN Ia rate are due to merging episodes with
small companion galaxies: the accretion of material both fuels the
central radio source and produces star formation, which boosts the SN Ia
production. In this scenario the observed excess implies that a
significant component of the SN Ia population is characterized by delay
times as short as those associated with the radio activity.

The lifetime of the radio emission after an episode of star formation in
not well known, but several works \citep{srinand98,wan00} constraint it
to be between 0.5 and $1.5\cdot10^8$ yr. After this period the radio
activity is expected to decay quickly. By assuming that the radio
activity and the new episode of star formation are coeval, the excess of
SN Ia is directly related to the integral of the DTD during the radio
lifetime. To calculate the SN rate associated with a given radio power,
we assume that, after a merging episode producing new stars for a few
percent of the total mass,  an early type galaxy is radio-loud for
$0.5-1.5\cdot10^8$ yr, radio-faint for a similar amount of time,
and radio-quiet afterward (see the definitions in
\citet{dellavalle05}).  As a result, the SN rate of the radio-loud and
radio-faint galaxies is expected to be proportional to the average of
the DTD during the first $\sim10^8$ and $\sim5\cdot10^8$ yr,
respectively, while the SN rate of the radio-quiet galaxies is
proportional to this average over one Hubble time. In other words, the
DTDs of SNe in the radio-loud and radio-faint galaxies must show an
excess at short times, otherwise  the rate distribution with radio power
would be flat.

\section{The predictions of the single-population DTDs}
\label{sec:single}

Fitting exclusively the redshift evolution of SN Ia rates
\citet{dahlen04} and \citet{strolger04}  (see also
\citet{strolger05}) derived indications for long delay times, i.e., a
DTD peaked at about 3.4 Gyr with basically no stars exploding
during the first two Gyr.  This result is the direct consequence of
the observation of a decreasing rate at z$>$1. We have tested if such a
DTD  can reproduce the other two observational evidences.  In this
case the only  parameter that can be varied to simultaneously reproduce
the observations is the explosion efficiency $\eta$.  The results  are
shown in Figure~1, where panel (a) shows the used DTD, panel (b) the
evolution of the SN rate with redshift, and panels (c) and (d) the
dependence of the SN rate with  the (B$-$K) colour and radio power of
the parent galaxies, respectively. It is apparent that such a DTD, while
accounting very well for the SN Ia rate evolution with redshift, is
totally inadequate to explain the two trends observed in the local
universe:  the statistical probability that the expectations are in
agreement with the observations are below 0.1\% for the rate-colour plot
and below 2.5\% for the rate-radio power plot.  This is true also when
considering the uncertainties in the model, shown as a grey area in
panel (c): in all cases the 
probability of agreement remains well
below 0.1\%. This demonstrates that the estimated delay times (about
3-4 Gyr) are far too long to reproduce the observed variation with radio
power and the enhancement of the rate in the blue galaxies: as no
SNe are expected to explode during the first couple of Gyr,  this DTD
cannot reproduce any dependence of the SN rate on the  current SFR.\\

Looking for a DTD that satisfies all the constraints, we have
investigated a large number of possible  ``single-population'' models,
i.e., DTDs that can be associated to a single progenitor population and
be described by a single analytical law. We used DTDs characterized by
different shapes (exponential decline, gaussian shape, and constant over
one Hubble time) and characteristic times between 0.1 and 6 Gyr. The
results are summarized in the first part of Table~1. None of these
simple DTDs can satisfy all of these observational constraints 
simultaneously. In all cases, at least one of the constraints is not
satisfied and the deviations are much greater than the observational
uncertainties.

Within this class of models, the observations are best
matched by an exponential distribution of delay times with e-folding
time of 3 Gyr (see Figure~2). This distribution provides a rather good
description of the observed rates as function of redshift (panel b) and
of the parent galaxy colours (panel c). However, this model is unable to
describe satisfactorily the variation of the rates with the radio-power
of the parent galaxy (panel d). 
These results suggests that while a DTD that
extends over several Gyr is needed, an additional 
contribution at early times (below $10^8$ yr), 
is necessary to explain all observations.

\begin{figure}
\centerline{\includegraphics[width=7.0cm]{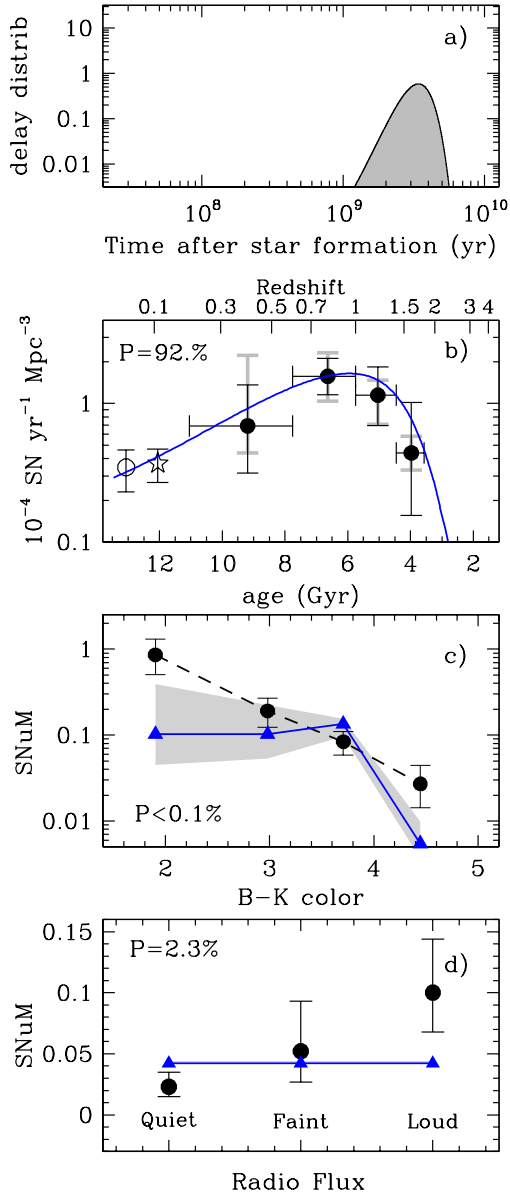} }
\caption{
Properties of the SN rates derived for a DTD having a gaussian
shape, centered in 3.4 Gyr and with $\sigma$=0.68 Gyr, as in
\citet{strolger05}. {\em Panel (a):} the DTD itself, plotted as number of
explosions per unit time after star formation; 
{\em Panel (b):} the evolution of the rate along the cosmic age, with
the upper axis showing the corresponding redshift. 
The solid line shows the results for the considered DTD.
Data are from \citet{mannucci05} (white circle),
\citet{strolger03} (starry dot), 
\citet{dahlen04} (black dots, 
with black error bars showing the 1$\sigma$ Poisson
statistical errors and the gray bars the systematic uncertainties
reported by the authors). 
P is the statistical probability of agreement derived 
from the $\chi^2$ values and takes into account the statistical errors only.
{\em Panel (c):} the predictions (solid line and triangles) of this 
DTD for the dependence of the rates on the parent galaxy (B$-$K) colour, 
expressed in SNe per century per $10^{10}$ \msun\ of stellar mass, SNuM.  
The gray region shows the peak-to-peak uncertainties in these 
predictions, as explained in the text. 
The dots and the dashed line show the observations in \citet{mannucci05}. 
{\em Panel (d):} variation of the
rates in the early-type galaxies with the radio power of the parent
galaxy. The black dots are the observed SN rates in \citet{dellavalle05}
with 1-$\sigma$ Poisson error.
}
\end{figure}

\begin{table}
\caption{List of some of the tested models together 
with the resulting values of $\chi^2_r$.
In boldface the values below 1.5. Many theoretical models providing poor
matching of the data are not listed}
\begin{tabular}{lcccc}
\hline
\hline
~~Model &$\eta$ &\multicolumn{3}{c}{$\chi^2_r$}\\
      & (\%)  &     z  &  colour  & radio\\
\hline
\multicolumn{4}{c}{Single population models} \\
\hline
Exponen. decay, $\tau$=1 Gyr        & 5.3 &2.7     & 7.2    & 3.7 \\ 
Exponen. decay, $\tau$=2 Gyr        & 4.0 &\bf{1.0}& 4.5    & 3.7 \\ 
Exponen. decay, $\tau$=3 Gyr        & 3.3 &\bf{0.8}& 4.5    & 3.7 \\ 
Exponen. decay, $\tau$=6 Gyr        & 3.2 &2.1     & 15.7   & 3.7 \\ 
Constant,                           & 2.5 &4.2     & 29     & 3.7 \\ 
Gauss at 0.05 Gyr, $\sigma$=0.01 Gyr& 6.5 &3.9     & 32     & 16  \\ 
Gauss at 0.5 Gyr,  $\sigma$=0.1 Gyr & 6.0 &4.0     & 37     & 3.7 \\ 
Gauss at 1 Gyr,    $\sigma$=0.2 Gyr & 5.4 &3.9     & 25     & 3.7 \\ 
Gauss at 1 Gyr,    $\sigma$=1.0 Gyr & 4.6 &2.4     & 9.2    & 3.7 \\ 
Gauss at 2 Gyr,    $\sigma$=0.4 Gyr & 4.0 &1.9     & 23     & 3.7 \\ 
Gauss at 2 Gyr,    $\sigma$=2.0 Gyr & 3.7 &\bf{0.5}& 15     & 3.7 \\ 
Gauss at 3.4 Gyr,  $\sigma$=0.68 Gyr& 3.6 &\bf{0.2}& 22     & 3.7 \\ 
Gauss at 4 Gyr,    $\sigma$=2.0 Gyr & 3.1 &\bf{1.0}& 14     & 3.7 \\ 
\hline
\multicolumn{4}{c}{Theoretical models} \\
\hline
YL00$^a$, DD-Ch,                    &3.8&\bf{1.0}&\bf{0.1}&2.4 \\ 
YL00, He-ELD,                       &4.5&2.2     &34      &3.2 \\
YL00, SG-Ch,                        &5.0&2.5     &21      &3.7 \\
YL00, SG-ELD,                       &4.0&\bf{0.8}&11      &3.7 \\
BBR05$^b$, SDS, \al=0.3             &4.0&\bf{0.7}&\bf{1.2}&1.7 \\
BBR05, SWB, \al=1.0                 &2.2&3.5     &13      &3.7  \\ 
G05$^c$ wide DD \tn=0.4 $\beta=-0.9$&4.0&\bf{1.1}&\bf{0.6}&3.1  \\
G05   close DD \tn=0.5 $\beta=-0.75$&4.0&\bf{1.0}&\bf{0.2}&2.9  \\
G05     SD chandra                  &4.1&\bf{0.9}&\bf{0.6}&3.3 \\
G05     SD sub-chandra              &4.3&\bf{1.2}&\bf{0.3}&3.0 \\
MR01$^d$  SD                        &4.7&\bf{1.5}&1.6     &2.2      \\  
\hline
\multicolumn{4}{c}{Two Populations models} \\
\hline
50\% prompt + 50\% gauss. 4 Gyr  & 4.3 &\bf{0.5}&5.0     &\bf{0.5}\\
50\% prompt + 50\% expon. 3 Gyr  & 4.3 &\bf{1.2}&\bf{0.7}&\bf{0.4}\\
50\% prompt + 50\% const.        & 3.0 &2.0     &6.2     &\bf{0.4}\\

\hline
\hline
\end{tabular}
$^a$ \citet{yungelson00};
$^b$ \citet{belczynski05};
$^c$ \citet{greggio05}
$^d$ \citet{matteucci01}
\end{table}

\begin{figure}
\centerline{\includegraphics[width=7.0cm]{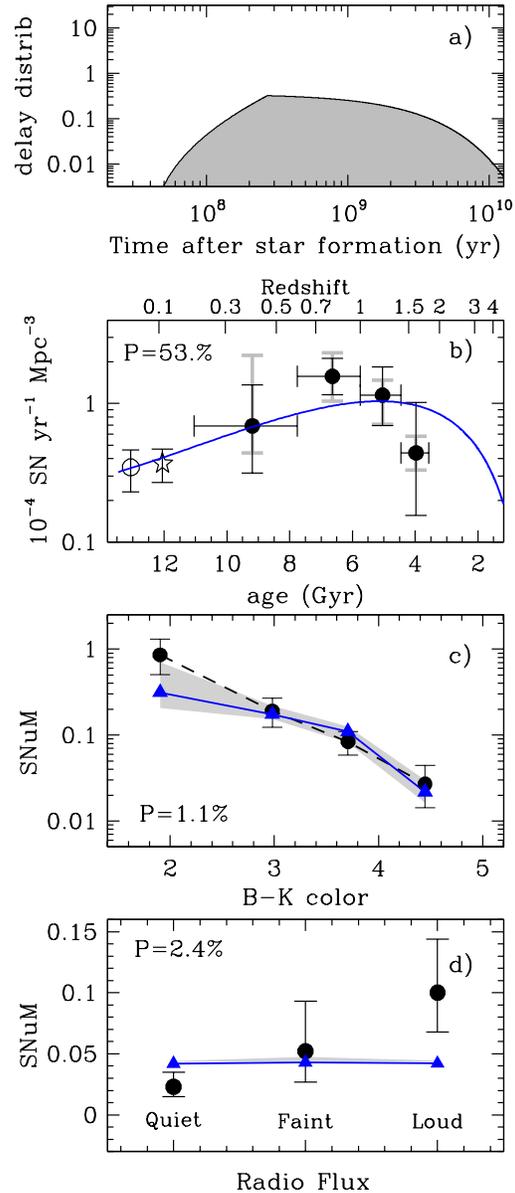} }
\caption{
Same as Figure~1, but  for a DTD that is an exponential function with an
e-folding time of 3 Gyr. The rise of DTD at about $10^8$ yr in panel (a)
is due to the building up of the white dwarf population.  It is apparent
that while the fit to the variation of the rates with redshift (panel b)
host galaxy colours (panel c) are  quite satisfactory when considering
the uncertainties of the models,   the behavior of the SN Ia rates with
radio luminosity in elliptical galaxies (panel d) cannot be reproduced.}
\end{figure}


\section{Theoretical models}
\label{sec:teo}

Several authors have derived the DTD from the models of stellar
evolution and SN explosion 
\citep{greggio83,yungelson00,matteucci01,belczynski05,greggio05}.
In Table~2 we report the results of the comparison between
``theory'' and ``observations'' by changing the only free parameter
$\eta$.  All models predicting a narrow DTD are ruled out by
observations.

Some of the single-degenerate and double-degenerate models, which predict very
broad DTDs, spanning delay times from $10^7$ yrs up to $10^{10}$ yrs,
give interesting results. This is the case for a number of
\citet{greggio05}  models, both SD and DD (see e.g. Figure~3), the
\citet{yungelson00} DD Chandrasekar mass model (see Figure~4), and the
the \citet{matteucci01} SD model shown in Figure~5. 
In all these models the DTD
peaks at about $0.6-2\cdot10^8$ yrs and then decays rapidly, 
roughly like $t^{-1}$, becoming ten
times lower after about 1 Gyr.  The dependence of the rates with
redshift and galaxy colours are satisfactory reproduced, even if 
in some cases the fast evolution tends to under-predict 
the SN rate in the reddest galaxies.
Most of these DTDs predict that about 5-15\% of the SNe explode
within the first $10^8$ yrs. As a consequence they produce a SN Ia rate
in radio loud galaxies only 10-40\% larger then in radio-quiet galaxies,
instead of the observed factor of 4.   Given the low number of observed
events, we cannot exclude these models with a 
very high statistical confidence level
(P$_{radio}$=2-10\%).
As an illustration,  let's consider 
\citet{greggio05} DD model
(Figure~3) that  well reproduces panels (b) and (c).
In this case the
peak of the DTD occurs at about $2\cdot10^8$ yrs and the fraction of SNe
in the first $10^8$ yrs is only 5\%. As a consequence the enhancement of
the rate in the radio-loud galaxies is only 10\%.   

Even if these
models cannot be ruled out with an high degree of confidence,
it is evident that a DTD with both more SNe at early
time and a slower  evolution afterward is needed to fully account for
the observations.

\begin{figure}
\centerline{\includegraphics[width=7.0cm]{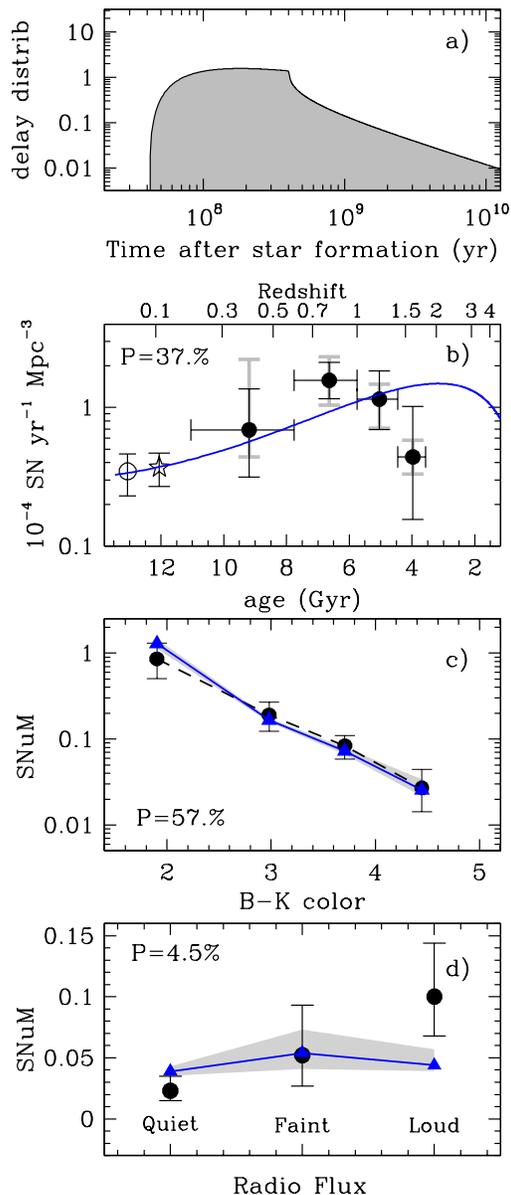}}
\caption{ 
Same as Figure~1, but for the theoretical DTD predicted
by \citet{greggio05} for her DD model for a wide binary.
The grey area in panel (d) corresponds to different assumptions for the 
lifetimes of the radio activity, from 0.5 to $1.5\cdot10^8$ yrs for the 
radio-loud and radio-faint phases, 
and to a fraction of new stars between 1 and 3\% in mass.
This DTD reproduces the observed trends in panels (b) and (c),
while the expected enhancement of the rate in radio-loud galaxies 
is much lower than observed factor of 4.}
\end{figure}

\begin{figure}
\centerline{\includegraphics[width=7.0cm]{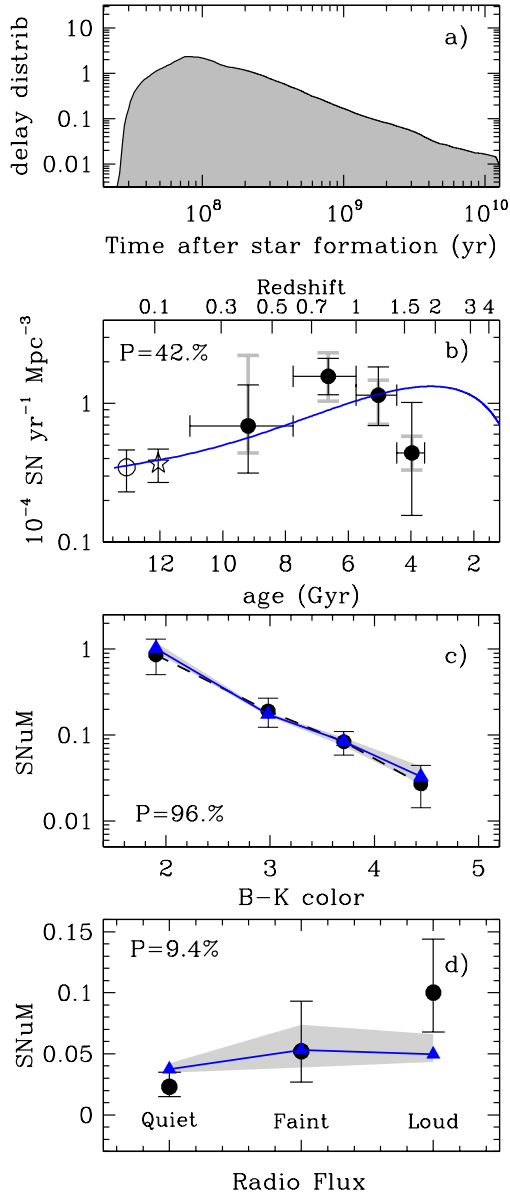}}
\caption{ 
Same as Figure~1, but for the DTD from the DD model
with Chandrasekar mass by \citet{yungelson00}.  
}
\end{figure}

\begin{figure}
\centerline{\includegraphics[width=7.0cm]{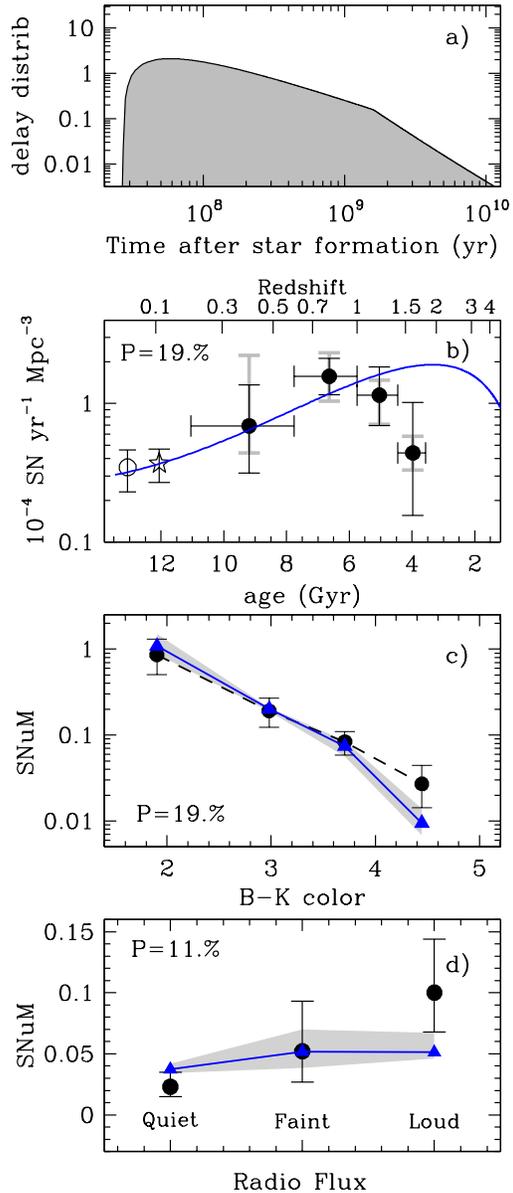}}
\caption{ 
Same as Figure~1, but for the theoretical DTD predicted
by the SD model in \citet{matteucci01}. 
}
\end{figure}

\begin{figure}
\centerline{\includegraphics[width=7.0cm]{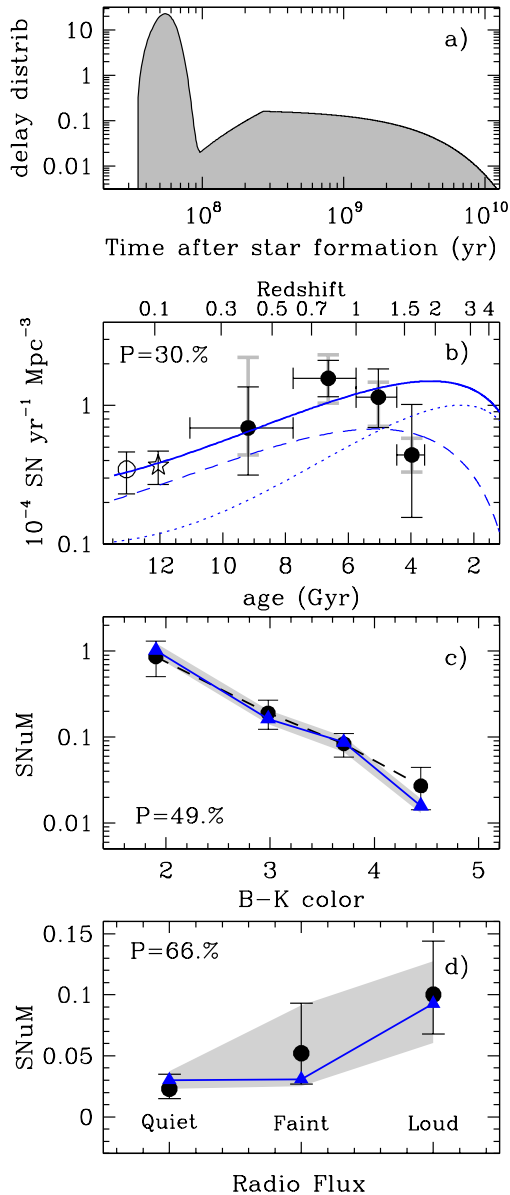}}
\caption{
Same as Figure~1, but for a DTD constituted by equally contributions
(50\%) of an exponentially declining function with characteristic time
of 3 Gyr  and a gaussian centered at $5\times10^7$ yr 
and $\sigma=10^7$ yr.
In panel (b)
the dotted and dashed lines show, respectively, the contributions from
the ``prompt'' and ``tardy'' components. 
The DTD simultaneously reproduces all
the SN Ia rate observations.
}
\end{figure}

\begin{figure}
\centerline{\includegraphics[width=7.0cm]{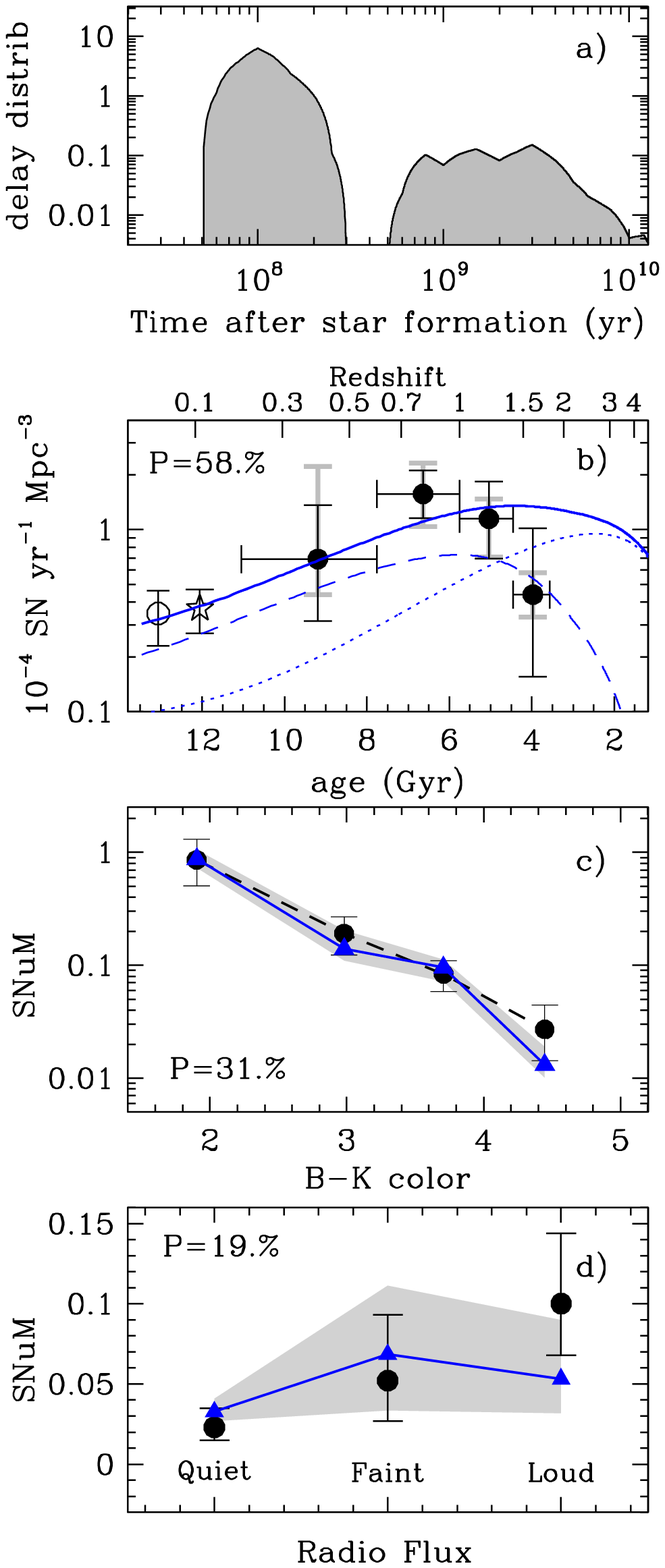}}
\caption{
Same as Figure~1, but for the theoretical DTD predicted by Belczynski et
al. (2005) for the single-degenerate model with reduced common envelope
efficiency ($\alpha\lambda=0.3$). The best-fitting
value of the total efficiency is $\eta$=4.0\%. In Panel (b) the dotted 
and dashed lines show, respectively, the contributions from the
``prompt'' and ``tardy'' components associated to the two peaks of the DTD. 
This bimodal DTD appears to reproduce
reasonably well the SN Ia rate observations,  although the 
position and the intensity of the ``prompt'' peak produce
only a small enhancement of the rate in the radio-loud Ellipticals.
}
\end{figure}

\section{Two populations DTDs}
\label{sec:twopop}

For these reasons we considered a set of ``two populations'' models in
which the DTD is obtained as the sum of two distinct functions. In all
cases, we added a ``prompt'' gaussian centered at $5\times10^7$ yr to a
much slower function, either another gaussian or an exponentially
declining function.  We will refer to the former component as ``prompt''
exploders and to latter as ``tardy'' ones\footnote{We chose to 
use ``tardy", rather than``delayed",
to avoid any misunderstanding with the
``delayed'' detonation model adopted for type Ia supernovae, e.g.
Woosley (1990), or more recently, Golombek \& Niemeyer (2005)}

Figure~4 shows the results of a model in which 50\% of the SNe derive
from the ``prompt'' population, and the remaining 50\% from the
``tardy'' one that consists in an exponentially declining function  with
characteristic time of 3 Gyr.  
These results shows that the observational data are better
reproduced by a DTD with a peak at short times  (below $10^8$ yr) that
includes about half of the SN Ia events, and an extension toward very
long times, say, 3 Gyr and beyond. Provided that the ``tardy'' component
extends well beyond 3 Gyr, its shape is not well  constrained by the
fit: exponential decays with characteristic times  between 2.5 and 8 Gyr
can still provide reasonable fits. These uncertainties will reduce
considerably when the SN Ia rates at z$>$0.5 will be measured with
greater accuracy.\\

We notice that a bimodal distribution of delay times should 
not be regarded as a simple heuristic method to better fit the data, 
because there are theoretical models which
actually predict a bimodal DTD.  One of
the best examples, see Figure~7, is obtained by \citet{belczynski05}
with a single-degenerate model with reduced common envelope efficiency
($\alpha\lambda=0.3$).  This model predicts a bimodal DTD which peaks at
$10^8$ and $3\cdot10^9$ yr, and is due to the explosions of both He and
C-O white dwarfs.  This model correctly reproduces the evolution of the
rate with redshift and its dependence on the colours, and it
accounts for the enhancement in the radio-loud galaxies only
qualitatively, because its ``prompt''  peak is centered at $10^8$ yr
instead at the best-fitting value of  $5\cdot10^7$ yr.  

Bimodal DTDs can be produced also by models with more than one type of
progenitors, for example in which both the single-degenerate and
double-degenerate channels are active (see, for example,
Nomoto et al., 2003).  A bimodal DTD is also naturally produced by the
SD model by \citet{koba98} in which two different companion stars are
present: either a red giant with initial mass of about 1 \msun\ and
orbital periods of tens to hundreds days,  or a main-sequence star with
mass $\sim2-3$ \msun\ and periods  of the order of a day.\\


\section{Discussion}

\subsection{Other evidences for bimodality}

We have shown that the observational constraints to the SN Ia
rates, namely the rate evolution with redshift, the dependence of SN Ia
rates with host galaxy colors, and the marked increase of SN Ia rates in
radio-loud Ellipticals, are best reproduced if about half of the SNe
explode within $10^8$ yr from star formation (``prompt'' component)
while the rest have explosion timescales of a few Gyr (``tardy''
component).  Particularly from our analysis a number of facts emerge:

\begin{itemize}

\item After constraining  the delay time distributions  to match both the 
observations at high redshift and in the local universe, we derive 
a DTD characterized by a strong peak at early times ($\sim$ 0.1 Mpc) 
followed by an exponential function with a decay time of about 3 Gyr. 
Our result rules out all models predicting narrow DTD,
pure exponential or pure gaussian DTD and constant DTD.

\item There are some single-population models, characterized by
broad distributions of delay times (3 orders of magnitude), which can
account the observations, but only at a modest statistical confidence
level. Given the current uncertainties on the SN rates, they cannot
be confidently ruled out. 

\item The best match to the data is obtained with a bimodal DTD.
However, the existence of a bimodality in the DTD does not directly
imply the existence of two distinct populations of SNe. This
bimodality might be related to the bimodal distribution of 
$\Delta m_{15}$ 
(the magnitude decline of the light curve between its
maximum and 15 days later) exhibited by SNe Ia in Ellipticals and in
spirals (Della Valle et al. 2005, their fig.~6, see also Altavilla et
al. 2004). If this is the case, the bimodality of the DTD gives
support to earlier suggestions which interpreted the bimodal behavior
of $\Delta m_{15}$ in terms of an age sequence (Ruiz-Lapuente et
al. 1995, Hamuy et al. 1996, Howell 2001, van den Berg et al. 2005),
with metallicity being less important (Ivanov et al. 2000, but see
Kobayashi et al. 1998). In this scenario  it is likely to expect that
the bright SN events (with small $\Delta m_{15}$) are preferentially
associated with the ``prompt'' population, which is more common in 
spiral and irregular galaxies (Hamuy et al. 2000), while the faint and
rapidly decaying objects are more easily found in the ``tardy''
component, which is more common in  early-type galaxies. If
this is true, then we expect that SN Ia in radio-loud Ellipticals,
where the ``prompt'' population is expected to dominate, show $\Delta
m_{15}$ smaller that those in the radio-quiet Ellipticals.  This issue
is discussed in \citet{dellavalle05} (see their fig.  6). 
The small number of events with published value of 
$\Delta m_{15}$ prevents us from reaching any definite conclusion.

\item Finally we notice that the existence of a continuous
sequence in the spectroscopic properties of maximum light spectra of
SNe Ia (Branch et al. 2006) is not at variance with the existence of two
different evolutionary paths to rise SNe Ia, as our results may suggest.
What is relevant for our analysis is how the spectroscopic properties of
SNe Ia correlate with the stellar population from which the SN
progenitors originate. The systematic differences measured in the
expansion velocities of the ejecta of SNe Ia occurring in Ellipticals
and Spirals (Branch \& van den Bergh 1993; Benetti et al. 2005) are
indeed fully consistent with our results.
\end{itemize}

\subsection{Bimodality and stellar mass}
\label{sec:mass}

We note that the main sequence lifetime is about $5\times10^8$ yr for
a star of 3\msun, $10^8$ yr for 5.5\msun\ and about $4\times10^7$ yr
for 8\msun\ (e.g., Girardi et al. 2000). Therefore, the SNe of the
``prompt'' peak, which include about 50\% of the total number of SN Ia
events and explode within $10^8$ yr from star formation, must all
derive from stars with masses above 5.5\msun. Also, for a Salpeter
IMF, the number of stars between 5.5 and 8\msun\ are about a third of
those between 3 and 5.5\msun. This implies that the SN Ia efficiency
for higher mass progenitors (5.5--8\msun) is about 3 times higher that
for lower mass progenitors (3--5.5\msun). Therefore, given an overall
efficiency of 4.5\% (see Figure~6), it follows that the efficiency for
higher mass stars is $\eta$(5.5--8\msun)$\sim$6.8\%, and the one for
lower mass stars is $\eta$(3--5.5\msun)$\sim$2.3\%.  Therefore,
the requirement that about 50\% of the SN Ia must explode within
$10^8$ yrs is that {\em the  efficiency and the
characteristic delay time are expected to change considerably
at $\sim$5.5\msun}.

It is important to realize that the explosion efficiency
of the ``prompt'' SN Ia is determined unambiguously by their
number and the mass range of the progenitors.
On the other hand, if the remaining 50\% SNIa, i.e.
the ``tardy"  component, could arise not only from stars in the range
3-5.5\msun,  but also from stars of lower masses, then their inferred
explosion efficiency for low mass stars would also decrease, because the
available pool of stars would increase whereas the number of ``tardy"
SNe does not.

We cannot draw conclusions on whether this change of the efficiency at
about $5.5$\msun\ is due to a different physical process (e.g., SD
vs. DD) or to one and the same process  operating in separate regions of
the parameter space (e.g., systematic differences of the binary
systems as a function of the stellar mass).  For example, the
theoretical models producing wide DTDs  described in
section~\ref{sec:teo}  \citep{yungelson00,belczynski05,greggio05} can be
modified to reproduce the empirical evidence if  this strong variation
of the efficiency with mass is introduced in  the models, boosting the
``prompt'' part of the DTD. Thus, it could be that the binary
fraction for primary stars with masses above 5.5\msun~ is markedly
higher than for lower mass stars. Or it could be that the distribution 
of secondary star masses is more skewed toward masses close to the
primary star mass and, therefore, the mass transfer be more efficient
and faster (see Pinsonneault \& Stanek 2006 for a discussion).

The existing models \citep{greggio05,belczynski05,nomoto03} are not able to
resolve this ambiguity because of both uncertainties in the model
assumptions and possible coexistence of different physical processes.
However, we are confident that a judicious analysis of data obtained for
a large sample of SN Ia over a suitably wide interval of redshifts will
make it possible to clarify this issue.

\subsection{Consequences of the bimodality on Cosmology}

In addition to providing essential clues to the nature of SN Ia
progenitors, our results have also important  implications for
cosmology:
\begin{itemize}
\item {\em The fraction of SNe coming from either of the two populations
changes with cosmic time}, as can be seen from Figure~6:  the ``tardy''
SNe dominate at z$<$1.3 and the ``prompt'' SNe  above this limit. The
ratio of the ``prompt'' SN rate to that of the ``tardy'' SNe changes
from 0.5 in the local universe to about 1.2 at z=1.5. Similar results
are obtained for the single-degenerate model by \citet{belczynski05}
(Figure~7).
\item  
It is reasonable to expect that the two populations of SNe can be
distinguished also by some intrinsic properties. As an example, it is
possible that ``prompt'' SNe Ia are, on average, more affected by dust
extinction than the ``tardy'' component, as they must explode closer  to
the formation cloud \citep{sullivan03}. In this case, {\em the average
properties of SNe Ia are expected to change with redshift}, especially
at z$>1$ when the ``prompt'' SNe become common.  The Hubble diagrams
used to derive information on the cosmological parameters
\citep{riess04} are, up to now, mostly based on SNe at z$<1$ and,
therefore, are expected  to be dominated by the ``tardy'' population. As
the ratio between the two different flavors of type SNe Ia changes with
cosmic time,  {\em evolutionary effects should become important at high
redshifts} (Livio \& Riess, 2006). 

\item The luminosity-decline rate relation for SN Ia
\citep{pskovskii77,phillips93,hamuy96,phillips97} is derived in the
local universe and, therefore, under this scenario,
is dominated by the ``tardy'' SNe.
The evidence for a cosmological acceleration relies on the
assumption that the same relation holds also at high redshift
(see Rowan-Robinson 2002 and Leibundgut 2004 for a discussion).
If the two populations follow slightly different relations, 
a bias is expected to emerge as a function of redshift,
especially when the ``prompt'' population becomes 
dominating, at $z\sim 1.2$. 
Thus, a confident use of
SNe Ia for cosmology measurements at z$>1$ would require a good
understanding of the differences in properties of the two populations.
\end{itemize}

\subsection{Bimodality and metallicity evolution}

The existence of two populations of SN Ia has direct consequences also
on the chemical evolution of the universe. 
\begin{itemize}
\item ``Prompt'' type Ia, having a redshift distribution similar to the
CC SNe, dominate the SN Ia population at high redshifts. Therefore, {\em
in the early Universe the production of Fe  is expected to follow that
of Oxygen}, and the O/Fe abundance ratio should be relatively constant
but appreciably higher than in the local Universe. When  the SN Ia
``tardy'' component starts dominating, i.e.  past the SFH peak at
z$\sim2$, the Fe production is boosted and the O/Fe ratio is expected to
decrease rapidly to approach the ``solar" values  around redshifts
$<\sim0.5$. These aspects have recently been discussed in some detail by
\citet{scanna05} who  for their model calculations  adopted the
simplified description of the SN Ia rates as derived by
\citet{mannucci05} in terms of a component proportional to the star
formation rate (SFR) and another one that is proportional to the  total
stellar mass.

\item It is known that the intra-cluster medium is relatively
rich in iron ([Fe/H]$\sim-0.5$) and that the metallicity shows a very
mild evolution with redshift (Tozzi et al., 2003). The observed iron
mass is about a factor of 6 larger than  could have been produced by
core-collapse SNe \citep{maoz04}  and a factor of 10 larger than that
produced by the current rate  of SN Ia (Renzini, 2004). The ``two
populations'' model naturally explains  these observations, as the
current type Ia rate is just the long-time  declining tail of a SN
distribution that peaked at early cosmic times. The amount of observed iron and
its redshift evolution is reproduced by assuming an average age of the 
stars in clusters of 10 Gyr (see Matteucci et al., in preparation).

\end{itemize}

\begin{figure}
\centerline{\includegraphics[width=7.9cm]{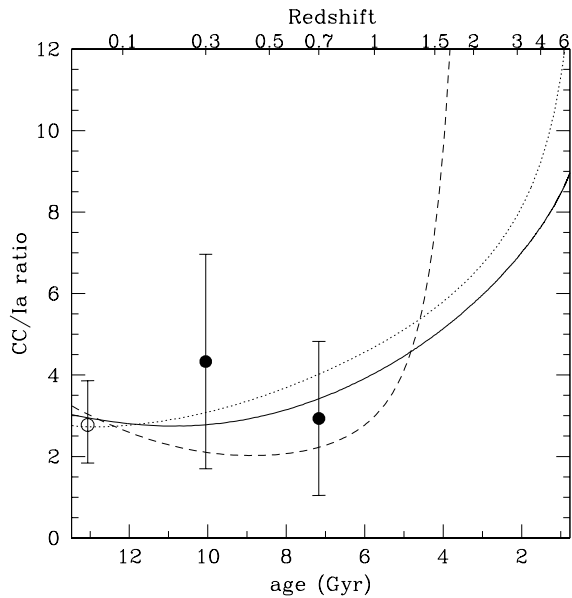} }
\caption{
Ratio of the rates of the CC to Ia SNe as a function of the redshift.
The white and black dots are observed values
\citep{mannucci05,dahlen04}.  The lines show the
predictions of the gaussian ``single-population'' model in figure~1 
(dashed), the \citet{yungelson00} model in figure~4 (dotted), and the
``two-populations'' in figure~6 (solid).  The predictions
use a Salpeter IMF  and mass ranges of 3-8\msun\ (SNe Ia) and 8-40\msun\
(CC SNe), and are scaled to match the observed values.
}
\end{figure}

\subsection{Predictions}

The existence of the ``prompt'' and ``tardy'' populations of type Ia SNe
can be tested by two observations:
\begin{itemize}
\item
The SN Ia rate is expected not to decrease significantly 
moving toward high redshifts up to z$\sim$2, 
at which the cosmic star formation history has its broad
peak.  As a consequence, {\em it should be possible to detect SNe Ia up
to high redshifts, z$\sim5$} so as to discriminate among  different
cosmological models.  In particular, at z$>$2 the SNe Ia rate should be
nearly constant at a level of about $10^{-4}$~SN~yr$^{-1}$~Mpc$^{-3}$
(see Figure~6).  Such a rate can be reduced only if the effects of
metallicity evolution become important at z$\sim$1, and if the changing
in metallicity has an important effect in the explosion rate as
predicted by \citet{koba98}.
\item  In the models predicting either bimodal or wide DTDs,
``prompt'' type Ia and Core-Collapse (CC) SNe are characterized
by similar delay times and both trace the cosmic star formation
history.  At high redshifts the SN Ia ``tardy'' component tends to
disappear and  therefore {\em we predict that the rate ratio CC/Ia
steadily increases  with redshift}, as shown in Figure~8, from a value
of about 3 in the local universe to about 9 at z$\geq$4. On the
contrary, the ``single-population'' model in Figure~1  predicts a much
faster evolution of the CC/Ia ratio, which is expected to become larger
than 10 already at z$\sim$1.5. 
\end{itemize} 
Measuring the SN rates and their CC/Ia ratio at high redshifts  will be a
very interesting task for the  upcoming James Webb Space Telescope
and giant ground-based telescopes 
and will permit to confirm or discard these predictions.\\

{ \bf Acknowledgments}
The authors are deeply indebted with D. Branch, L. Greggio, F.
Matteucci, and S. Recchi  for useful discussions, and to K. Belczynski
and L. Yungelson for providing the results of their models in a digital
format. The  authors also wish to thank an anonymous referee for
comments that have helped to  improve the presentation of this paper.


\end{document}